\documentclass[10pt,twoside,twocolumn,british,nofootinbib, aps]{revtex4}
\usepackage{color}

\usepackage{amsmath}
\usepackage{graphicx}
\usepackage{amssymb}
\usepackage{esint}
\usepackage[unicode=true, 
 bookmarks=true,bookmarksnumbered=false,bookmarksopen=false,
 breaklinks=false,pdfborder={0 0 1},backref=false,colorlinks=true]
 {hyperref}

\makeatletter
\@ifundefined{textcolor}{}
{%
 \definecolor{BLACK}{gray}{0}
 \definecolor{WHITE}{gray}{1}
 \definecolor{RED}{rgb}{1,0,0}
 \definecolor{GREEN}{rgb}{0,1,0}
 \definecolor{BLUE}{rgb}{0,0,1}
 \definecolor{CYAN}{cmyk}{1,0,0,0}
 \definecolor{MAGENTA}{cmyk}{0,1,0,0}
 \definecolor{YELLOW}{cmyk}{0,0,1,0}
 }

\usepackage{hyperref}
\hypersetup{
    colorlinks,%
    citecolor=blue,%
    filecolor=blue,%
    linkcolor=blue,%
    urlcolor=blue
}

\makeatother

\begin{document}

\title{Dust of Dark Energy}

\author{Eugene A. Lim$^{1}$}

\email{eugene.a.lim@gmail.com}

\author{Ignacy Sawicki$^{2}$ }

\email{ignacy.sawicki@nyu.edu}

\author{Alexander Vikman$^{2}$}

\email{alexander.vikman@nyu.edu}

\affiliation{$^{1}$ISCAP and Department of Physics, Columbia University, New
York, NY 10027, USA}

\affiliation{$^{2}$CCPP, Department of Physics, New York University, New York,
NY 10003, USA}

\date{\today{}}
\begin{abstract}
We introduce a novel class of field theories where energy always flows
along timelike geodesics, mimicking in that respect dust, yet which
possess non-zero pressure. This theory comprises two scalar fields,
one of which is a Lagrange multiplier enforcing a constraint between
the other's field value and derivative. We show that this system possesses
no wave-like modes but retains a single dynamical degree of freedom.
Thus, the sound speed is always identically zero on all backgrounds.
In particular, cosmological perturbations reproduce the standard behaviour
for hydrodynamics in the limit of vanishing sound speed. Using all these properties
we propose a model unifying Dark Matter and Dark Energy in a single
degree of freedom. In a certain limit this model exactly
reproduces the evolution history of $\Lambda$CDM, while deviations
away from the standard expansion history produce a potentially measurable
difference in the evolution of structure.
\end{abstract}
\maketitle

\section{Dusty Fluid with Pressure?}

How can one obtain dust from a scalar field? One can imagine a canonical
scalar-field where the kinetic term is constrained to be equal to
the potential. We can implement this property by introducing a Lagrange
multiplier, $\lambda$, in the Lagrangian,

\[
\mathcal{L}=\lambda\left(\frac{1}{2}(\partial\varphi)^{2}-V(\varphi)\right)\,.\]
We then find that the pressure is identically vanishing on all solutions
and energy follows geodesics. This model describes the usual dust
without vorticity. 

How can we obtain ``dust with pressure''? We can generalise the above
by adding some function of the scalar field and its derivatives to
the Lagrangian,\[
\mathcal{L}=K\left(\varphi,\partial\varphi\right)+\lambda\left(\frac{1}{2}(\partial\varphi)^{2}-V(\varphi)\right)\,.\]
The constraint remains in effect and standard scalar-field dynamics
are not restored. In fact, we will show that fluid elements in all
such theories \emph{also }always\emph{ }flow along geodesics, mimicking
in that respect standard dust, yet the fluid has non-vanishing pressure.
With this simple idea we have separated the notion that the pressure
of the fluid is tied to the motion of a fluid element as is the situation
in the usual case, e.g. radiation or cold dark matter. A parcel of
such fluid will flow along geodesics, yet a manometer will record
a pressure changing with time. 

In this paper, we introduce this new class of scalar-field models,
which we will call $\lambda\varphi$-\emph{fluids}. These theories
are described by an action containing two scalar fields, $\varphi$
and $\lambda$, where the latter plays the role of a Lagrange multiplier
and enforces a constraint relating the value of the scalar field $\varphi$
to the norm of its derivative. This constraint forces the dynamics
of the $\lambda\varphi$-\emph{fluid} to be driven by a system of
two first-order \emph{ordinary }differential equations, one for the
field $\varphi$, the other for the Lagrange multiplier. As a consequence,
there are no propagating wave-like degrees of freedom and the sound
speed for perturbations is exactly zero irrespective of the background
solution. However, the initial-value problem still requires the specification
of two functions on the initial time slice. Thus, effectively, a single
dynamical degree of freedom remains.

Provided that the derivatives of the scalar field $\varphi$ are time-like,
the system can be interpreted as a perfect fluid. However, for a $\lambda\varphi$-\emph{fluid}
given by a particular action, the relation between the pressure $p$
and the energy density $\varepsilon$ is solution dependent. We show
that an arbitrary effective equation of state, including phantom ones,
can be obtained by choosing the form of the Lagrangian appropriately.
In addition, for all $\lambda\varphi$-\emph{fluids}, there always
exists a region in their phase spaces in which the $\lambda\varphi$\emph{-fluid}
is effectively pressureless. We will exploit this feature to model
the evolution of the cosmological background from matter domination
through to the acceleration era as being driven by the dynamics of
a single degree of freedom provided by the $\lambda\varphi$\emph{-fluid}. 

The key novel aspect of this class of theories is that the four-acceleration
is always zero, even when the pressure does not vanish. This is a
result of the constraint's eliminating those fluid configurations
where the pressure has a gradient orthogonal to the fluid velocity. 

Motivated by these properties, we will use the $\lambda\varphi$\emph{-fluid}
to frame the problem of the dark sector in cosmology in a unified
manner. The existence of the dark sector in the Universe's energy
budget is now established beyond reasonable doubt. The standard model
of cosmology, $\Lambda$CDM, splits it into two constituents: cold
dark matter (``CDM'')---a pressureless fluid (``dust'') which clusters
allowing baryonic structures to form in its potential wells and detectable
to this day in the form of halos around galaxies---and dark energy
(``DE'')---a form of energy that appears to be smooth and to have
an equation of state close to a cosmological constant. This dichotomy
of phenomenology has made it difficult to build a compelling model
which would treat the two dark components in a unified fashion. Nonetheless,
some models exist in the literature: \cite{Padmanabhan:2002sh, Scherrer:2004au,Bertacca:2007ux, Arbey:2006it, Kamenshchik:2001cp, Bento:2004uh, Capozziello:2005tf, Nojiri:2005pu, Quercellini:2007ht, Linder:2008ya, Piattella:2009kt, Gao:2009me,Creminelli:2009mu}. It is not hard to see that, given a fluid which
clusters like dust yet has arbitrary pressure, we can construct such
a unified model---which we call Dusty Dark Energy (``DDE''). We present
the main results of our application of $\lambda\varphi$\emph{-fluid}
to such a cosmology beneath. We refer the reader to the full analysis
in the main body of the paper, section \ref{s:DDE}.

\subsection{Summary of Cosmological Results}

We have studied cosmological perturbations in the case when an arbitrary
$\lambda\varphi$\emph{-fluid} dominates the\emph{ }Universe. We have
derived the closed-form equation for the evolution of the Newtonian
potential $\Phi$ which turns out to recover the standard result for
general hydrodynamics in the limit of vanishing sound speed. This
evolution is determined by background expansion history only and in
the limit of the $\Lambda$CDM expansion history the evolution of
perturbations is exactly as in the standard case. We have also written
down an action for perturbations which explicitly shows that there
are no ghosts in this theory when the equation of state for the $\lambda\varphi$-\emph{fluid}
is non-phantom. We demonstrate that on a classical level our model
can cross the phantom divide without singularities while linear perturbations
continue to evolve stably; however, the perturbations do become ghosts
at this point, hence making the system unstable (in particular quantum-mechanically)
when interactions are taken into account. In this paper we put the
investigation of that instability and issues related to the possible
strong coupling scales aside. 

We consider a universe comprising only radiation and the $\lambda\varphi$-\emph{fluid}
which will describe both CDM and DE. To illustrate more concretely
the phenomenology we have focused on a specific family of models which
is parameterised by $w_{\text{fin}}$---the equation of state of the
$\lambda\varphi$-\emph{fluid} in the asymptotic future. Given that
the initial values for the $\lambda\varphi$-\emph{fluid} are chosen
appropriately, the radiation becomes subdominant while the $\lambda\varphi$-\emph{fluid}
is still far off its final attractor (given by $w_{\text{fin}}$)
and evolves approximately like dust, giving an epoch of matter domination.
The duration of this epoch is also determined by the initial values.

\begin{figure}
\includegraphics[width=1\columnwidth]{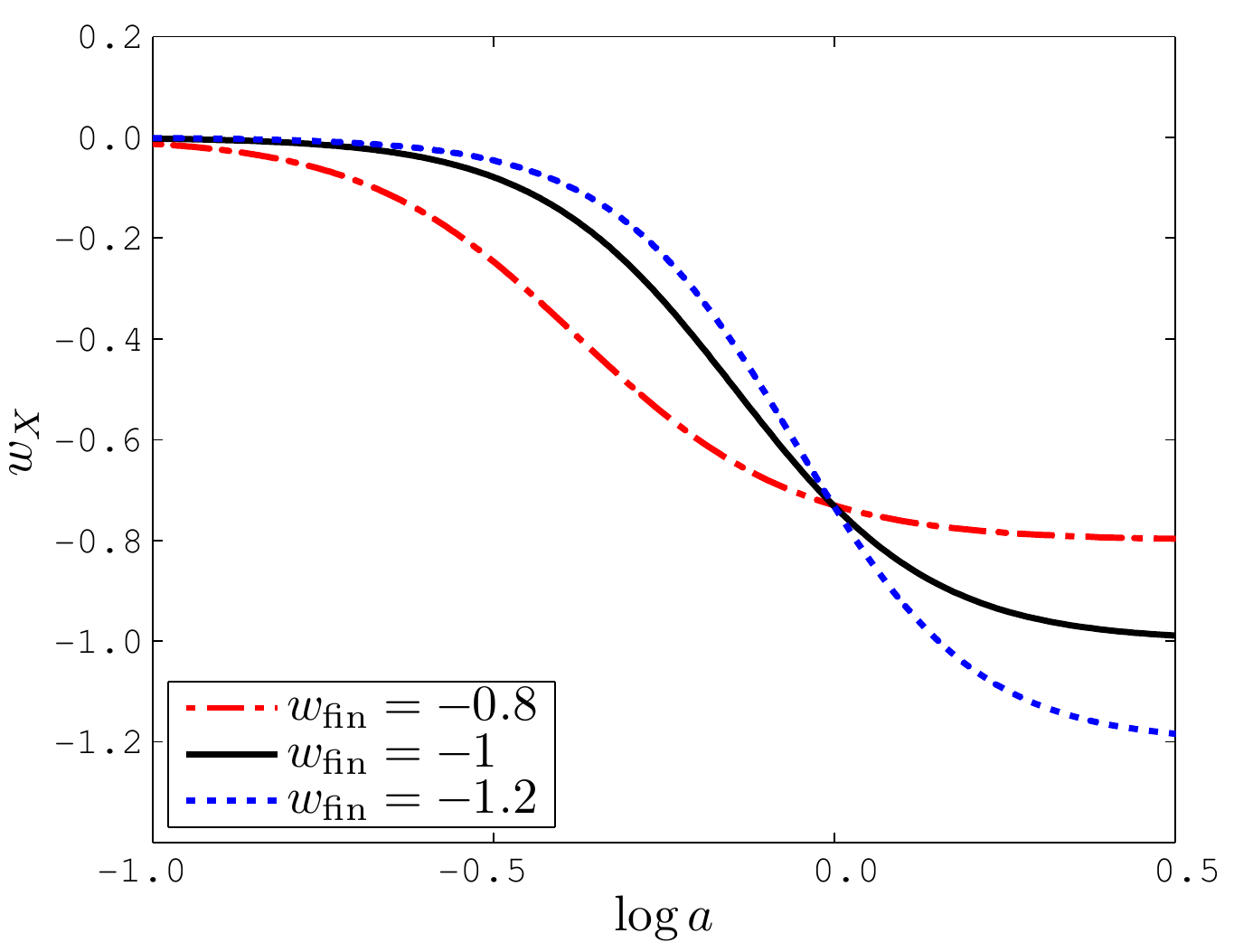}

\caption{Time evolution of the total effective equation of state for the dark
sector. The black solid line represents the evolution in $\Lambda$CDM
which is identical to that of the $w_{\text{fin}}=-1$ model. Models
with final equations of state $1+w_{\text{fin}}>0$ begin to deviate
from matter domination earlier and the transition is slower than $\Lambda$CDM.
The opposite is true in phantom models. The evolution is normalised
such that the equation of state at $a=1$ matches the best-fit result
for the $\Lambda$CDM cosmology as determined by WMAP7 results, $w_{0}=-0.74$
\cite{Komatsu:2010fb}. \label{fig:wevol} }

\end{figure}

\begin{figure}
\includegraphics[width=1\columnwidth]{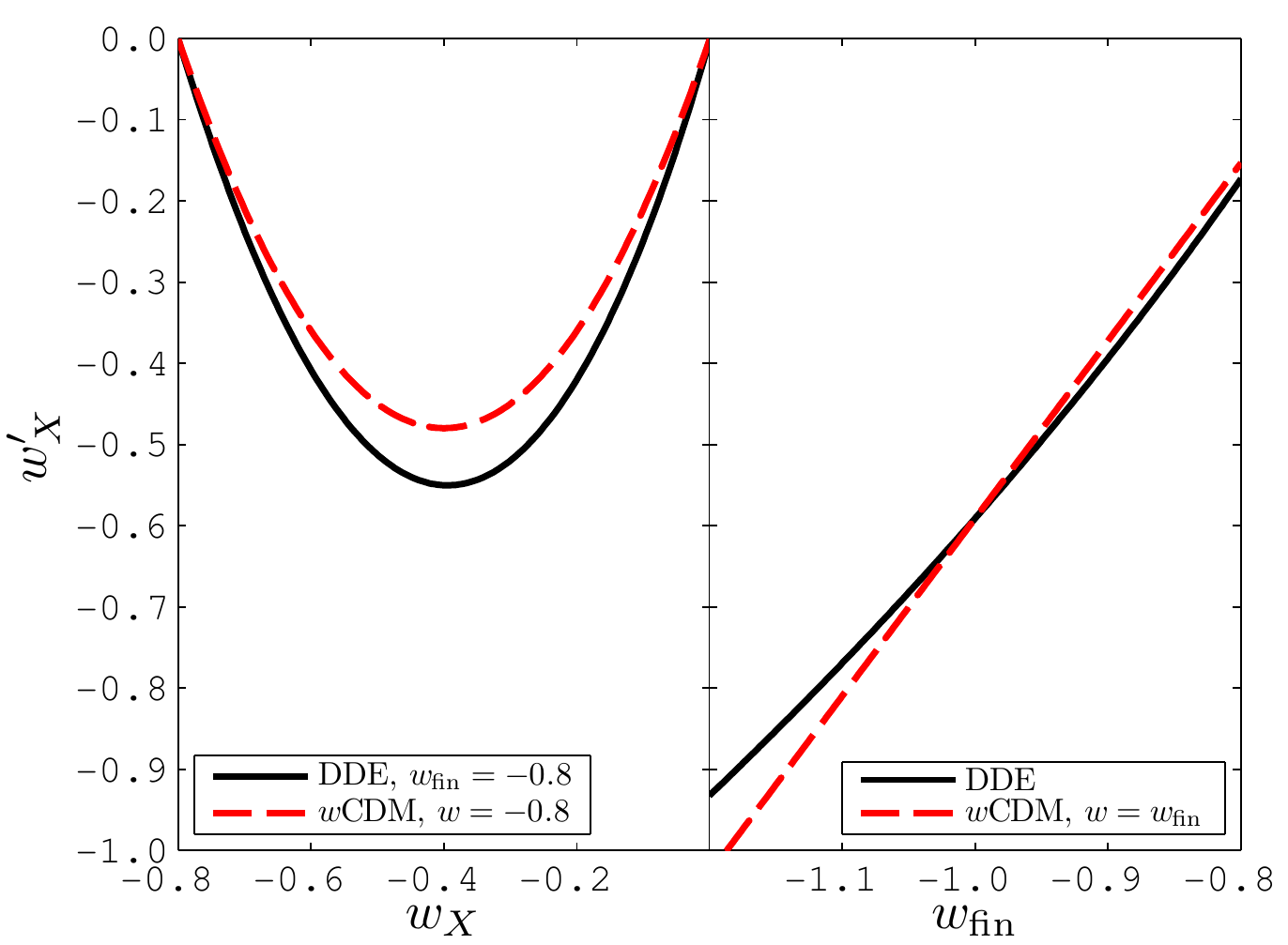}\caption{Comparison of the derivative of the effective equation of state for
Dusty Dark Energy (Eq.\ \eqref{LambdaEvolutionWeqConst}) with that
for a dark matter plus dark energy with a constant equation of state,
$w$CDM, (Eq.\ \eqref{WevolutionMixture}). The magnitude of the derivative
determines the duration of the transition between matter domination
and the acceleration era. The left panel shows that for $w_{\text{fin}}>-1$,
the transition in the DDE model is more rapid than the corresponding
$w$CDM model. On the other hand, for phantom $w_{\text{fin}}$ this
transition is slower than for the corresponding $w$CDM model, as
shown in the panel on the right. \label{fig:wprime}}

\end{figure}

\begin{figure}
\includegraphics[width=1\columnwidth]{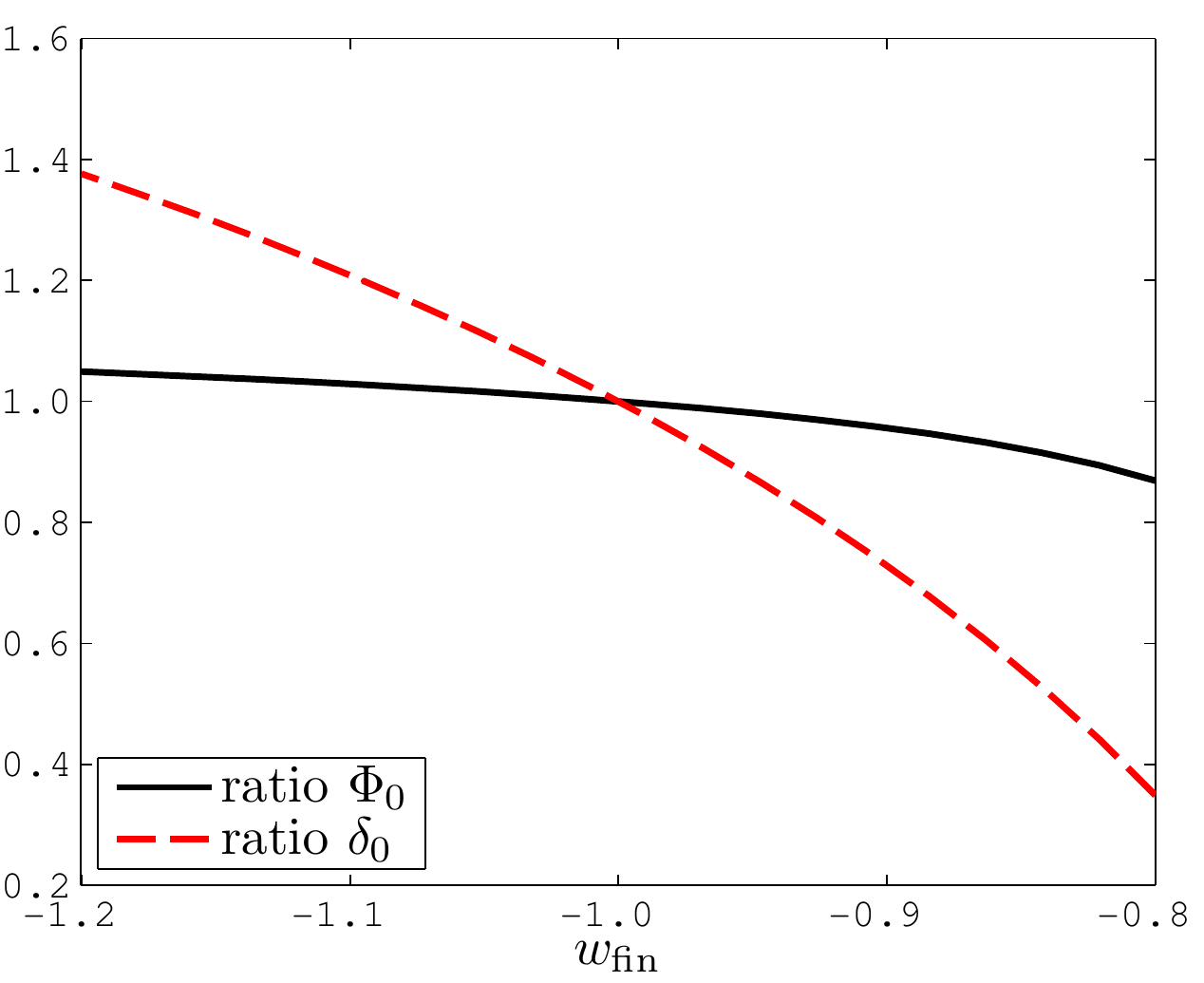}\caption{The comparison of the total growth of perturbation amplitude between
Dusty Dark Energy and $\Lambda$CDM. The evolution of the Newtonian
potential, $\Phi$, is determined by Eq.\ \eqref{PhiN} and deviates
by a few percent from its $\Lambda$CDM values by a few percent. This
evolution will affect the strength of the ISW signal in the CMB. On
the other hand, the evolution of the density perturbation on subhorizon
scales is determined by Eq.\ \eqref{dotRelativeDeltaE} and is affected
much more strongly.\textbf{\label{fig:Ratio}}}

\end{figure}

In the limit $w_{\text{fin}}\rightarrow-1$ this family of models
recovers exactly the background evolution and growth of structure
of $\Lambda$CDM. However, if $w_{\text{fin}}\neq-1,$ the evolution
of the background and perturbations differs from a ``$w$CDM'' model
comprising cold dark matter and dark energy with a constant equation
of state. We illustrate the evolution of the effective equation of
state for the dark sector in a selection of different $w_{\text{fin}}$-cosmologies
in Fig.\ \ref{fig:wevol}. We find that the transition from matter-domination
to dark energy domination differs from that of $w$CDM (Fig.\ \ref{fig:wprime}).
We compare the growth of linear perturbations with that of $\Lambda$CDM
in Fig.\ \ref{fig:Ratio}. We find that models with $1+w_{\text{fin}}>0$
exhibit a growth factor suppressed by a few tens of percent. Also
the Newtonian potential here is lower by a few percent than in $\Lambda$CDM,
which would decrease the integrated Sachs-Wolfe effect. 

\subsection{Open Questions and Future Directions}

Finally, we mention open issues in this setup which remain to be addressed 
\begin{itemize}
\item The nature of caustics: It is well known that in non-canonical field
theories caustic can develop, e.$\,$g. see Ref. \cite{Felder:2002sv}
on caustics in Sen's string-theoretical tachyon matter \cite{Sen:2002in,Sen:2002nu}
and Ref. \cite{ArkaniHamed:2005gu} on caustics in the ghost condensate
\cite{ArkaniHamed:2003uy} and the discussion in Refs \cite{Mukohyama:2009tp,Blas:2009yd}
on caustics in Ho\v{r}ava gravity \cite{Horava:2009uw}. We expect
that the $\lambda\varphi$\emph{-fluid }will develop caustics. The
question is how to interpret the multivalued regions once this occurs.
\item Can $\lambda\varphi$\emph{-fluids} virialise? If the $\lambda\varphi$\emph{-fluid}
is to really model non-linear structure, it must be able to form static
and stable configurations (e.g. halos). 
\item What is the origin for the initial conditions for Dusty Dark Energy
introduced in section \ref{s:DDE}? Could a more generalised setup
provide a solution for the coincidence problem? In our model, we require
that at some point during the radiation epoch the energy density of
the $\lambda\varphi$\emph{-fluid} be equal to the energy density
of CDM: we have no alternative to the standard dark-matter freeze
out scenario which would produce this in a natural manner.
\item What is the Hamiltonian structure of this theory? How to quantise
it? See, for example \cite{Brown:1994py}. Is the structure of this theory stable as a result of radiative corrections: would a kinetic term for $\lambda$ be generated?
\item What is the strong-coupling scale for perturbations? We should note
that our model is rather similar to a potentially singular limit of
Ho\v{r}ava gravity \cite{Horava:2009uw}. There it was found \cite{Blas:2009yd}
that, contrary to our case, the sound speed for cosmological perturbations
becomes imaginary, and that the strong coupling scale may be extremely
low.
\item What is the rate of instability for the phantom case? Is this instability
catastrophic as it is usually for ghost degrees of freedom? It is
possible that the absence of propagating wave-like degrees of freedom
may change the standard picture \cite{Carroll:2003st,Dubovsky:2005xd,Woodard:2006nt,Cline:2003gs,Emparan:2005gg}.
\item The action formulation of this theory allows us to consider couplings
to standard model fields. It is natural, for example, to consider
Lorentz violation in this framework in analogy to Einstein aether
theories \cite{Jacobson:2004ts,Carroll:2004ai,Lim:2004js}.
\item Can this theory be a low-energy limit of a more fundamental theory? 
\item Can our conceit with the constraint be usefully extended beyond scalar
fields to theories with fermions, vector fields or many degrees of
freedom?
\end{itemize}

\section{Dynamics in General Curved Space-Time}

Let us consider a scalar field theory given by the action \[
S=\int\mbox{d}^{4}x\sqrt{-g}\left(K\left(\varphi,X\right)+\lambda\left(X-\frac{1}{2}\mu^{2}\left(\varphi\right)\right)\right)\,,\]
where the field $\lambda$ is a {}``Lagrange multiplier'' and does
not have a kinetic term, while \[
X\equiv\frac{1}{2}g^{\alpha\beta}\nabla_{\alpha}\varphi\nabla_{\beta}\varphi\,,\]
 is a standard kinetic term for the field $\varphi$, $K\left(\varphi,X\right)$
is an arbitrary function of $X$ and $\varphi$, while $\mu\left(\varphi\right)$
is an arbitrary function of the scalar field $\varphi$. In hydrodynamical language, $\varphi$ is one of the velocity potentials and does not necessarily have to represent a fundamental degree of freedom; $\mu(\varphi)$ plays the role of the specific inertial mass \cite{Schutz70}. Let us further
assume a standard minimal coupling with gravity. 

The equations of motion are \begin{eqnarray}
 &  & \!\!\!\!\frac{1}{\sqrt{-g}}\frac{\delta S}{\delta\lambda}=X-\frac{1}{2}\mu^{2}\left(\varphi\right)=0\,,\label{EOMLambda}\\
 &  & \!\!\!\!\frac{1}{\sqrt{-g}}\frac{\delta S}{\delta\varphi}=K_{\varphi}-\nabla_{\alpha}\left(K_{X}\nabla^{\alpha}\varphi\right)-\label{EOMPhi}\\
 &  & \!\!\!\!\qquad\qquad\qquad\qquad-\lambda\mu\mu_{\varphi}-\nabla_{\alpha}\left(\lambda\nabla^{\alpha}\varphi\right)=0\,.\nonumber \end{eqnarray}
Here and throughout the paper we denote partial derivatives by subscripts.
The Lagrange multiplier imposes a non-holonomic constraint%
\footnote{As a technical point, the system possesses two second class constraints,
one which is primary and the other secondary.%
} on the dynamics of the fields, hence the action cannot be written
purely in terms of $\varphi$. In the case of time-like derivatives:
$X>0$, similarly to k-\emph{essence} \cite{ArmendarizPicon:1999rj,ArmendarizPicon:2000ah,ArmendarizPicon:2000dh},
we can introduce an effective 4-velocity \begin{equation}
u_{\alpha}=\frac{\nabla_{\alpha}\varphi}{\sqrt{2X}}=\mu^{-1}\nabla_{\alpha}\varphi\,,\label{4U}\end{equation}
where in the last equality we have used the ``constraint'' Eq.\ (\ref{EOMLambda}).
Further, we can see that the corresponding effective four-acceleration
always vanishes, \begin{equation}
a_{\beta}=\dot{u}_{\beta}=\left(\mu^{-1}\nabla_{\gamma}\varphi\right)\nabla^{\gamma}\left(\mu^{-1}\nabla_{\beta}\varphi\right)=0\,.\label{NoAcceleration}\end{equation}
Here and throughout the paper we use the notation $\dot{\left(\;\;\right)}=u^{\alpha}\nabla_{\alpha}\left(\;\;\right)$
for the derivative along $u^{\alpha}$. Thus, on equations of motion,
$u^{\alpha}$ are tangents to time-like geodesics. It is also convenient
to write Eq.\ (\ref{EOMLambda}) in this form \begin{equation}
\dot{\varphi}=\mu\left(\varphi\right)\,.\label{PhiEvolution}\end{equation}
The general solution of Eq.\ (\ref{PhiEvolution}) is \begin{equation}
\varphi\left(x^{\alpha}\right)=f\left(\tau-\psi\left(\mathbf{x}\right)\right)\,,\label{phiTau}\end{equation}
where $\tau$ is a time parameterising the congruence of geodesics
given by $u^{\alpha}$, $f$ is a general function solving Eq.\ (\ref{PhiEvolution})
and $\psi\left(\mathbf{x}\right)$ is an arbitrary function of spatial
coordinates $\mathbf{x}$ in the hypersurface normal to $u^{\alpha}$.
One can consider $\varphi$ to be an \emph{intrinsic clock} for our
system, since Eq.\ \eqref{phiTau} defines a time reparameterization. 

The energy-momentum tensor (EMT) is \[
T_{\alpha\beta}=\frac{2}{\sqrt{-g}}\frac{\delta S}{\,\,\delta g^{\alpha\beta}}=\left(K_{X}+\lambda\right)\nabla_{\alpha}\varphi\nabla_{\beta}\varphi-Kg_{\alpha\beta}\,,\]
which is of the perfect-fluid form. Using hydrodynamical notation
the EMT can be rewritten as \[
T_{\alpha\beta}=\left(\varepsilon+p\right)u_{\alpha}u_{\beta}-pg_{\alpha\beta}\,,\]
with the energy density given by \begin{equation}
\varepsilon\left(\lambda,\varphi\right)=\mu^{2}\left(K_{X}+\lambda\right)-K\,,\label{EnergyDensity}\end{equation}
where into $K_{X}$ and $K$ we substitute the constraint $X=\frac{1}{2}\mu^{2}$.
The pressure is a function of the intrinsic clock\emph{ }$\varphi$
only \begin{equation}
p\left(\varphi\right)=K\left(\varphi,\frac{\mu^{2}\left(\varphi\right)}{2}\right)\,.\label{pressure}\end{equation}
This is the key feature responsible for the absence of acceleration,
Eq.\,\eqref{NoAcceleration}, since the gradient of the pressure
is always parallel to $u^{\alpha}$. This can be explicitly seen by
considering the conservation of the EMT: \begin{align*}
\nabla_{\alpha}T^{\alpha\beta} & =\left(\dot{\varepsilon}+\dot{p}+\theta\left(\varepsilon+p\right)\right)u^{\beta}+\left(\varepsilon+p\right)a^{\beta}-\nabla^{\beta}p=\\
 & =\left(\dot{\varepsilon}+\theta\left(\varepsilon+p\right)\right)u^{\beta}\,,\end{align*}
where we have used the fact that pressure does not have gradients
orthogonal to $u^{\alpha}$ and $\theta\equiv\nabla_{\alpha}u^{\alpha}$
is the expansion of a congruence of geodesics under consideration.
For perfect fluids in the usual case, the divergence of the EMT has
two components: one parallel to the velocity $u^{\beta}$ describing
the conservation of energy and one parallel to the four-acceleration
$a^{\beta}$. Here the latter is identically zero and so the conservation
of the EMT reduces to the conservation of energy,\begin{equation}
\dot{\varepsilon}+\theta\left(\varepsilon+p\right)=0\,.\label{EnergyConservation}\end{equation}
Then by choosing $K$ and $\mu$ appropriately one can arrange for
a general evolving equation of state \[
w_{X}\equiv\frac{p}{\varepsilon}\,.\]
Despite this, the energy flux of the $\lambda\varphi$\emph{-fluid,}
$T^{\alpha\beta}u_{\beta}=\varepsilon u^{\alpha}$, always follows
time-like geodesics, as is the case for perfect fluids in the absence
of pressure. Note that in FRW universes, the energy of an arbitrary
fluid flows along time-like geodesics as well, provided that the fluid
configuration be bound to respect the homogeneity and isotropy of
the FRW spacetime. This is of course not true of inhomogeneous perturbations
in the standard case. 

Now, let us refocus on the equation of motion. Using the constraint
Eq.\ (\ref{EOMLambda}) we can re-express the gradient of the kinetic
term and the expansion in terms of $\varphi$, \begin{align}
\nabla_{\alpha}X & =\mu\mu_{\varphi}\nabla_{\alpha}\varphi\,,\nonumber \\
\theta & =\nabla_{\alpha}u^{\alpha}=\mu^{-1}\Box\varphi-\mu_{\varphi\,}.\label{expansion}\end{align}
The system of equations of motion Eq.\ (\ref{EOMPhi}) and Eq.\ (\ref{EOMLambda})
can be written as\begin{align}
\dot{\varphi} & =\mu\left(\varphi\right)\,,\label{dotPhi}\\
\dot{\lambda} & =-\mu^{-2}\left(\varepsilon_{\varphi}\mu+\left(\varepsilon+p\right)\theta\right)\,,\label{dotlambda}\end{align}
where for the partial derivative of $\varepsilon_{\varphi}$ we differentiate
Eq.\ (\ref{EnergyDensity}) to obtain\begin{equation}
\varepsilon_{\varphi}=\mu\mu_{\varphi}\left(\mu^{2}K_{XX}+K_{X}+2\lambda\right)+\mu^{2}K_{X\varphi}-K_{\varphi}\,.\label{e_phi}\end{equation}
The equations of motion, Eqs (\ref{dotPhi}) and (\ref{dotlambda}),
for our system have been reduced in this way to two first-order ordinary
differential equations. They then have to be supplemented by the Landau-Raychaudhuri
equation \[
\dot{\theta}=-\frac{1}{3}\theta^{2}-\sigma_{\alpha\beta}\sigma^{\alpha\beta}-R_{a\beta}u^{\alpha}u^{\beta}\,,\]
for the expansion $\theta$ and similar equations for the shear $\sigma_{\alpha\beta}$
or, equivalently, by Einstein's equations%
\footnote{Note that the rotation $\omega_{\alpha\beta}$ is zero because of
Eq.\ (\ref{4U}).%
}. 

From the above system of equations of motion and the 4-velocity Eq.\
(\ref{4U}) it follows that the Cauchy problem locally has a unique
solution depending on two functions $\varphi\left(\mathbf{x}\right)$
and $\lambda\left(\mathbf{x}\right)$ given on an initial spacelike
hypersurface $\Sigma$. Moreover, from Eq.\ (\ref{dotPhi}) and Eq.\ 
(\ref{dotlambda}) it follows that the solutions propagate along time-like
geodesics. Therefore in the future of the geodesic $\gamma_{i}$ starting
at $\mathbf{x}_{i}$ on $\Sigma$ the solution only depends on the
initial values of $\varphi\left(\mathbf{x}_{i}\right)$ and $\lambda\left(\mathbf{x}_{i}\right)$
at the point $\mathbf{x}_{i}$---the sound speed $c_{\text{s}}$ is
identically equal to zero for all configurations of $\lambda\varphi$\emph{-fluid}.
This means that there are no wave-like dynamical degrees of freedom.
Adding dynamical gravity does not change the picture. 

As we have already mentioned in the introduction, the global Cauchy
problem may be ill-defined for some initial data since caustics may
develop. However, this problem is not unusual for non-canonical field
theories.

Finally we note that, since $\varepsilon_{\lambda}=\mu^{2}\neq0$,
we could use the pair $\left(\varphi,\,\varepsilon\right)$ instead
of $\left(\varphi,\,\lambda\right)$ as independent variables. In
that case, one has to use energy conservation Eq.\ (\ref{EnergyConservation})
instead of Eq.\ (\ref{dotlambda}) as the second equation of motion.

\section{Dynamics in Cosmology}

\subsection{Cosmological Background}

In the case of a background cosmology of a pure $\lambda\varphi$\emph{-fluid}
in the FRW universe with the metric \[
\mbox{d}s^{2}=\mbox{d}t^{2}-a^{2}\left(t\right)\mbox{d}\mathbf{x}^{2}\,,\]
our equations of motion take the form \begin{align}
\dot{\varphi}= & \mu\left(\varphi\right)\,,\label{CosmologyEoM}\\
\dot{\lambda}= & -\mu^{-2}\left(\varepsilon_{\varphi}\mu+3H\left(\varepsilon+p\right)\right)\,,\nonumber \end{align}
where we have used $\theta=3H\equiv3\dot{a}/a$. Alternatively, we
can use energy conservation Eq.\ \eqref{EnergyConservation} instead
of the equation for $\lambda$. To close this system we use the Friedmann
equation%
\footnote{Here, for simplicity, we have assumed a spatially flat case.%
}\begin{equation}
H^{2}=\frac{\varepsilon}{3M_{\text{Pl}}^{2}}=\frac{\mu^{2}\left(K_{X}+\lambda\right)-K}{3M_{\text{Pl}}^{2}}\,,\label{Friedmann}\end{equation}
where $M_{\text{Pl}}\equiv\left(8\pi G_{\mathrm{N}}\right)^{-1/2}$
is the reduced Planck mass. The phase space for this system of two
first-order ordinary differential equations is $\left(\lambda,\varphi\right)$. 

For completeness we also present the second Friedmann equation\begin{equation}
\dot{H}=-\frac{\varepsilon+p}{2M_{\text{Pl}}^{2}}=-\frac{\mu^{2}\left(K_{X}+\lambda\right)}{2M_{\text{Pl}}^{2}}\,.\label{SecondFriedmann}\end{equation}
The presence of an external energy density of some other cosmological
fluid brings with it the corresponding change to the Friedmann equations,
Eqs \eqref{Friedmann}, \eqref{SecondFriedmann} but no changes to
the equations of motion apart from those implied by the change in
$H$.

\subsection{Cosmological Perturbations }

\label{s:genperts}Symmetry considerations imply that our scalar field
at linear order only contributes to the scalar part of the cosmological
perturbations. In Newtonian gauge, the metric for scalar perturbations
is \[
\mbox{d}s^{2}=\left(1+2\Phi\right)\mbox{d}t^{2}-a^{2}\left(t\right)\left(1-2\Psi\right)\mbox{d}\mathbf{x}^{2}\,.\]
The absence of anisotropic stress, $\Phi=\Psi$, simplifies the perturbed
Einstein equations to (see e.g. \cite{Mukhanov:2005sc}): \begin{align}
 & \frac{\Delta}{a^{2}}\Phi-3H\left(\dot{\Phi}+H\Phi\right)=\frac{\delta\varepsilon_{\text{tot}}}{2M_{\text{Pl}}^{2}}\,,\label{deltaE}\\
 & \left(\dot{\Phi}+H\Phi\right)_{,i}=\frac{\left(\varepsilon+p\right)}{2M_{\text{Pl}}^{2}}\delta u_{\text{tot}\Vert i}\,,\label{deltaU}\\
 & \ddot{\Phi}+4H\dot{\Phi}+\left(2\dot{H}+3H^{2}\right)\Phi=\frac{\delta p_{\text{tot}}}{2M_{\text{Pl}}^{2}}\,,\label{deltaP}\end{align}
where $\Delta=\partial_{i}\partial_{i}$ and the $\delta u_{\text{tot}\Vert i}$
is the potential (scalar) part of the $i$ component of the total
four-velocity. Perturbing the equations of motion gives: \begin{equation}
\delta\dot{\varphi}=\mu_{\varphi}\delta\varphi+\Phi\mu\,,\label{PerturbtLambdaEOM}\end{equation}
for the constraint Eq.\ (\ref{EOMLambda}) and 

\begin{equation}
\delta\dot{\varepsilon}-\left(3\dot{\Phi}+\frac{\Delta}{a^{2}}\left(\frac{\delta\varphi}{\mu}\right)\right)\left(\varepsilon+p\right)+3H\left(\delta\varepsilon+\delta p\right)=0\,,\label{DotdeltaE}\end{equation}
for the perturbations of the energy density. See Appendix \ref{Appendix}
for the derivation. This equation is standard for hydrodynamical matter
\cite[p. 312, Eq.\ (7.105)]{Mukhanov:2005sc}. The nonstandard input
of our model is that $\delta p=p_{\varphi}\delta\varphi$ and that
Eq.\ \eqref{PerturbtLambdaEOM} describes the evolution of the velocity
potential $v=\mu^{-1}\delta\varphi$ for time-like geodesics. Note
that in the above equations, we have not assumed the domination of
the $\lambda\varphi$\emph{-fluid}. Thus, equations (\ref{PerturbtLambdaEOM})
and (\ref{DotdeltaE}) can be used to follow the dynamics of the linear
perturbations of $\lambda\varphi$\emph{-fluid} through the entire
history of the universe. For future discussion it is helpful to re-express
the above as an equation for the relative perturbation $\delta_{\varepsilon}=\delta\varepsilon/\varepsilon$:\begin{align}
\dot{\delta}_{\varepsilon}= & -3H\frac{p_{\varphi}}{\varepsilon}\delta\varphi+\label{dotRelativeDeltaE}\\
 & +3Hw_{X}\delta_{\varepsilon}+\left(3\dot{\Phi}+\frac{\Delta}{a^{2}}\left(\frac{\delta\varphi}{\mu}\right)\right)\left(1+w_{X}\right)\,.\nonumber \end{align}
Combining the above with the perturbed constraint Eq.\ \eqref{PerturbtLambdaEOM}
and obtaining the Newtonian potential $\Phi$ through the perturbed
Einstein equations \eqref{deltaE},$\,$\eqref{deltaU},$\,$\eqref{deltaP}
closes the system and allows us to describe the evolution of the perturbations
in the $\lambda\varphi$\emph{-fluid} in general case.

When the $\lambda\varphi$\emph{-fluid} dominates the energy density
and the perturbations, a significant simplification occurs since\begin{align*}
\delta u_{\text{tot}\Vert i}= & \mu^{-1}\partial_{i}\delta\varphi\,,\\
\delta\varepsilon_{\text{tot}}= & \varepsilon_{\varphi}\delta\varphi+\mu^{2}\delta\lambda\,,\\
\delta p_{\text{tot}}= & p_{\varphi}\delta\varphi\,.\end{align*}
Using these expressions for perturbations, we can first integrate
Eq.\ (\ref{deltaU}) to obtain\[
\delta\varphi=\frac{2M_{\text{Pl}}^{2}\mu}{\varepsilon+p}\left(\dot{\Phi}+H\Phi\right)=-\frac{\mu}{\dot{H}}\left(\dot{\Phi}+H\Phi\right)\,,\]
where in the last equality we have used the second Friedmann equation
\eqref{SecondFriedmann}. Combining this with Eq.\ \eqref{deltaP},
we can obtain a closed expression for $\Phi$\begin{align}
\ddot{\Phi} & +\dot{\Phi}H\left(4+\frac{\mu p_{\varphi}}{2M_{\text{Pl}}^{2}\dot{H}H}\right)+\label{closedEqForPotential}\\
 & +\left(2\frac{\dot{H}}{H^{2}}+3+\frac{\mu p_{\varphi}}{2M_{\text{Pl}}^{2}\dot{H}H}\right)H^{2}\Phi=0\,.\nonumber \end{align}
Thus, the Newtonian potential always evolves as \begin{equation}
\Phi\left(t,\mathbf{x}\right)=f_{1}\left(t\right)C_{1}\left(\mathbf{x}\right)+f_{2}\left(t\right)C_{2}\left(\mathbf{x}\right)\,,\label{solutionForPhi}\end{equation}
where $C_{1}\left(\mathbf{x}\right)$, $C_{2}\left(\mathbf{x}\right)$
are arbitrary functions of the spatial coordinates while $f_{1}\left(t\right)$
and $f_{2}\left(t\right)$ are solutions of the homogeneous ordinary
differential equation Eq.\ (\ref{closedEqForPotential}). Using the
solution Eq.\ (\ref{solutionForPhi}) we can find all other quantities.
In particular, substituting this solution for $\Phi$ into the Poisson
equation (\ref{deltaE}), we obtain $\delta\varepsilon$. The separation
of variables in the solution (\ref{solutionForPhi}) implies that,
as we have discussed above, the sound speed is zero. 

It is convenient to introduce dimensionless time (e-folds number)
$N\equiv\ln a$ and rewrite the differential equation for the Newtonian
potential $\Phi$ in terms of $N$\begin{equation}
\Phi''+\Phi'\left(4+\frac{H'}{H}+\Gamma\right)+\left(3+2\frac{H'}{H}+\Gamma\right)\Phi=0\,,\label{PhiN}\end{equation}
where we have introduced a dimensionless correction to the equation
for standard pressureless dust arising from the perturbation of pressure
in our model, \begin{equation}
\Gamma\equiv\frac{\mu p_{\varphi}}{2M_{\text{Pl}}^{2}H'H^{2}}\,,\label{GammaDEF}\end{equation}
with $\left(\;\right)'=\partial_{N}\left(\;\right)$ the derivative
with respect to the number of e-folds. Using the Friedmann equations,
$\Gamma$ can be written in geometrical terms as \[
\Gamma=-\frac{H''}{H'}-\frac{H'}{H}-3\,.\]
This particular combination vanishes for a background expansion history
mimicking that of $\Lambda$CDM. Therefore, in that limit, linear
perturbations in our model will behave identically to dust in the
presence of a cosmological constant.

Further, similarly to the standard case one can introduce a new variable
\[
Q=\sqrt{\frac{a}{-H'}}\Phi\,,\]
such that Eq.\ (\ref{PhiN}) takes the form of an oscillator with time-dependent
frequency\begin{eqnarray}
Q''-\left(\frac{\Theta''}{\Theta}\right)Q=0\,, & \mbox{where} & \Theta\equiv\frac{H}{\sqrt{-aH'}}\,.\label{Oscillator-1}\end{eqnarray}
Our variable $Q$ is a redefinition $Q\propto u\sqrt{aH}$ of the
standard variable \begin{equation}
u\propto\frac{\Phi}{\sqrt{-\dot{H}}}\propto\frac{\Phi}{\sqrt{\varepsilon+p}}\,,\label{uVariablePerturbations}\end{equation}
 given in \cite[p. 302, Eq.\ (7.63)]{Mukhanov:2005sc}. This redefinition
is caused by our choice of the time coordinate. The equation of motion
for $u$ is \begin{equation}
\partial_{\eta}^{2}u-\left(\frac{\partial_{\eta}^{2}\theta}{\theta}\right)u=0\,,\label{uEvol}\end{equation}
where $\eta$ is conformal time ($ad\eta=dt$) and \begin{equation}
\theta\propto a^{-1}\left(1+w_{X}\right)^{-1/2}\propto\frac{1}{a\sqrt{-\dot{H}}}\propto\frac{\Theta}{\sqrt{aH}}\,.\label{TETA}\end{equation}
It is easy to check that equations \eqref{Oscillator-1} and \eqref{uEvol}
are equivalent. Thus, cosmological perturbations of the $\lambda\varphi$-\emph{fluid
}reproduce the standard result for hydrodynamical matter in the limit
$c_{\mathrm{s}}=0$. We could have guessed the equation of motion
\eqref{uEvol} and the variable $u$ from the very beginning because
neither the final expression for $u$ nor the formula for $\theta$
explicitly involve the sound speed. However, note that the derivation
of these results presented in \cite[p. 302]{Mukhanov:2005sc} uses
in an essential way the standard hydrodynamical formula for the sound
speed $c_{\mathrm{s}}^{2}=\dot{p}/\dot{\varepsilon}$ which is clearly
absolutely inapplicable for the $\lambda\varphi$-\emph{fluid. }Curiously,\emph{
}$u$ and $\theta$ are given by the same formulae \eqref{uEvol}
and \eqref{TETA} in the case of k-\emph{essence} where the derivation
uses a sound speed given by $c_{\mathrm{s}}^{2}=p_{X}/\varepsilon_{X}\neq\dot{p}/\dot{\varepsilon}$
(see \cite{Garriga:1999vw,Mukhanov:2005sc}). 

From equation \eqref{Oscillator-1}, one can see that one of its solutions
is $Q\propto\Theta$, which translates to one of the modes for the
Newtonian potential \[
\Phi\propto\frac{H}{a}\,.\]
Taking the derivative gives the {}``instantaneous power law'': \[
\left(\ln\frac{H}{a}\right)'=-\left(\frac{5+3w_{X}}{2}\right)\,,\]
demonstrating that for $w_{X}>-5/3$ this solution represents the
decaying mode. Following the discussion in \cite[p. 303]{Mukhanov:2005sc},
we find the second mode of the solution from the Wronskian. Integrating
this result by parts allows us to write the full solution as\begin{equation}
\Phi=C_{1}\left(\mathbf{x}\right)+\frac{H}{a}C_{2}\left(\mathbf{x}\right)-C_{1}\left(\mathbf{x}\right)\frac{H}{a}\int^{a}\frac{\mbox{d}a}{H}\,.\label{PhiGenSoln}\end{equation}
This solution is valid on all scales, whereas in the standard case
when $c_{\mathrm{s}}\neq0$ it is only applicable on superhorizon
modes.

\subsubsection{Perturbations Around Scaling Solutions}

Let us analyse this general solution further in the case of a cosmology
dominated by a $\lambda\varphi$-\emph{fluid }with a constant equation-of-state
parameter $w_{X}\neq-1$ (we demonstrate how to construct one such
class of models in section \ref{s:constW}). In such a case, the final
term in the expression above is constant, and our solution for $w_{X}\neq-5/3$
is \begin{equation}
\Phi=\tilde{C}_{1}\left(\mathbf{x}\right)+C_{2}\left(\mathbf{x}\right)a^{-\left(5+3w_{X}\right)/2}\,,\label{PhiWconst}\end{equation}
while for the special case of $w_{X}=-5/3$ the solution is \[
\Phi_{-5/3}=\tilde{C}_{1}\left(\mathbf{x}\right)+\tilde{C}_{2}\left(\mathbf{x}\right)\ln a\,.\]
Thus, for all constant $w_{X}>-5/3$ the Newtonian potential is constant
up to a decaying mode. Substituting this solution \eqref{PhiWconst}
into the Poisson equation (\ref{deltaE}), we obtain for the density
fluctuations in the cases $w_{X}\neq-1$ and $w_{X}\neq-5/3$: \begin{eqnarray*}
\delta_{\varepsilon} & = & -2\Phi+\frac{2}{3}\frac{\Delta}{\left(aH\right)^{2}}\Phi-2\Phi'=\\
 & = & -2\tilde{C}_{1}+3\tilde{C_{2}}\left(1+w_{X}\right)a^{-\left(5+3w_{X}\right)/2}+\\
 &  & +\frac{2}{3}\frac{\Delta}{\left(aH\right)^{2}}\left(\tilde{C}_{1}+\tilde{C}_{2}a^{-\left(5+3w_{X}\right)/2}\right)\,.\end{eqnarray*}
For subhorizon modes, $k\gg aH$, we have\begin{align*}
 & \left(\delta_{\varepsilon}\right)_{k\gg aH}\sim a^{1+3w_{X}} &  & \mbox{for} & w_{X}>-\frac{5}{3}\,,\\
 & \left(\delta_{\varepsilon}\right)_{k\gg aH}\sim a^{3\left(w_{X}-1\right)/2} &  & \text{for} & w_{X}<-\frac{5}{3}\,,\end{align*}
while for superhorizon modes, $k\ll aH$, we have\begin{align*}
 & \left(\delta_{\varepsilon}\right)_{k\ll aH}\sim\text{const} & \mbox{for} &  & w_{X}>-\frac{5}{3}\,,\\
 & \left(\delta_{\varepsilon}\right)_{k\ll aH}\sim a^{-\left(5+3w_{X}\right)/2} & \mbox{for} &  & w_{X}<-\frac{5}{3}\,.\end{align*}
It is illustrative to compare these results with those for standard
cosmological fluids. In particular, for the ultrarelativistic equation
of state, $w_{X}=1/3$, the Newtonian potential for the $\lambda\varphi$\emph{-fluid}
remains constant up to a decaying mode at all scales, while subhorizon
density perturbations grow as $\delta_{\varepsilon}\sim a^{2}$. This
should be contrasted with standard radiation which causes both the
potential and the density perturbation to decay and oscillate once
the mode becomes subhorizon.

\subsubsection{Phantom Behaviour and $w=-1$ Crossing}

As the general solution \eqref{PhiGenSoln} implies, the perturbations
do not have catastrophic instabilities even for phantom \cite{Caldwell:1999ew}
equations of state, with $w_{X}<-1$. The $\lambda\varphi$\emph{-fluid}
framework allows one to realise such scenarios easily. However, Eq.
\eqref{GammaDEF} implies that $\Gamma$ has a singularity when $H'=0$,
i.e. when $w_{X}=-1$. Can a $\lambda\varphi$\emph{-fluid} evolve
through this singularity, from a standard to a phantom equation of
state?

The answer is provided by the analysis of Eq.\ \eqref{PhiN} in the
vicinity of the singularity. The derivatives of $\Phi$ can only remain
finite, if the singular terms in the equation cancel, i.e. if $\Phi'+\Phi=0$
at the phantom-divide-crossing point. Indeed, it can easily be checked
that the general solution \eqref{PhiGenSoln} always satisfies\[
\Phi'+\Phi=\frac{H'}{a}\left(C_{2}\left(\mathbf{x}\right)-C_{1}\left(\mathbf{x}\right)\int^{a}\frac{\mbox{d}a}{H}\right)\,,\]
so that automatically $\Phi'+\Phi=0$ when $H'=0$. Therefore, we
have also shown that in our model there are no classical catastrophic
instabilities associated with the crossing of the phantom divide. 

Note that the $\lambda\varphi$\emph{-fluid} contains only one degree
of freedom but the action cannot be written exclusively in terms of
this degree of freedom in a generally covariant local form. Therefore
this possibility of smooth crossing of the $w_{X}=-1$ barrier does
not contradict to the statement proved in \cite{Vikman:2004dc} and
rederived in different ways later in \cite{Caldwell:2005ai,Hu:2004kh,Zhao:2005vj,Sen:2005ra,Abramo:2005be,Kunz:2006wc,Babichev:2007dw}.
The $\lambda\varphi$\emph{-fluid} provides a working example of the
so-called \emph{Quintom} scenario from \cite{Feng:2004ad}, see also
reviews \cite{Cai:2009zp,Zhang:2009dw,Copeland:2006wr}. Further,
abandoning another assumption from \cite{Vikman:2004dc}, that coupling
to gravity is minimal, allows one to have a classically stable crossing
of the phantom divide in scalar-tensor theories \cite{Boisseau:2000pr,Gannouji:2006jm,Hu:2007nk,Motohashi:2010tb,Park:2010cw}. For other single-field options see \cite{Creminelli:2006xe, Creminelli:2008wc, Deffayet}. For more on phantoms see e.g.\ \cite{Nojiri:2005sr, Nojiri:2005pu, Capozziello:2005pa} and references therein.

One can obtain the equation of motion \eqref{uEvol} from the action
\begin{equation}
S_{u}=\frac{1}{2}\int\mbox{d}\eta\mbox{d}^{3}x\left(\left(\partial_{\eta}u\right)^{2}+\left(\frac{\partial_{\eta}^{2}\theta}{\theta}\right)u^{2}\right)\,.\label{Su}\end{equation}
Both quantities $u$ and $\theta$ (or $Q$ and $\Theta$) are defined
up to a constant factor. In particular, this factor can be a complex
number e.$\,$g. the imaginary unit $i$. Note that the definitions
\eqref{uVariablePerturbations}, \eqref{TETA} which we have used
imply that $u$ and $\theta$ are real, provided that the Null Energy
Condition for the background holds i.$\,$e. $\dot{H}<0$. As usual,
the sign of the action \eqref{Su} is such that for $\dot{H}<0$ it
has a positive definite kinetic term. When $\dot{H}=0$, both quantities
$u$ and $\theta$ diverge. However, as we have already demonstrated,
the evolution of the physical quantity $\Phi$ is smooth through $\dot{H}=0$.
This also means that the curvature invariants are smooth through $w_{X}=-1$
crossing. Thus after the crossing $u$ and $\theta$ become pure imaginary
or in terms of real fields, the action changes the overall sign and
the kinetic term becomes negative definite.

\section{Example $\lambda\varphi$-Fluid Cosmologies}

\subsection{$\lambda\varphi$-Dust with Cosmological Constant}

Let us consider the simplest case: $K_{\varphi}=0$ and $\mu=\mbox{const}$.
In that case, we have \begin{eqnarray*}
p\left(\mu\right)=K & \mbox{and} & \varepsilon=\mu^{2}\left(K_{X}+\lambda\right)-K\,,\end{eqnarray*}
where $K\left(\mu\right)=\mbox{const}$ and $K_{X}\left(\mu\right)=\text{const}$.
For the equations of motion, Eq.\ (\ref{CosmologyEoM}), we have \begin{align*}
\varphi= & \mu t\,,\\
\dot{\lambda}= & -3H\left(K_{X}+\lambda\right)\,,\end{align*}
with the solution \[
\left(K_{X}+\lambda\right)=\frac{\mu^{-2}\varepsilon_{0}}{a^{3}}\,,\]
where $\varepsilon_{0}$ is a constant of integration. Thus the system
behaves as\begin{eqnarray*}
p=\mbox{const} & \mbox{and} & \varepsilon=\varepsilon_{0}a^{-3}-p\,,\end{eqnarray*}
so that the energy-momentum corresponds to a mixture of a cosmological
constant $\Lambda=-K\left(\mu\right)$ of either sign and pressureless
dust with the energy density $\varepsilon_{0}$ today. Since the background
is that of $\Lambda$CDM cosmology, by the argument of section \ref{s:genperts},
the evolution of $\Phi$ and $\delta_{\varepsilon}$ also exactly
reproduces the standard results.

\subsection{$\lambda\varphi$-Dust}

If we take $K=0$ and an arbitrary $\mu\left(\varphi\right)$ we obtain
a $\lambda\varphi$-fluid which mimics pure dust, $p=0$, with energy
density $\varepsilon=\mu^{2}\lambda$. Indeed from Eq.\ (\ref{CosmologyEoM})
we have \begin{align*}
\dot{\varphi}= & \mu\left(\varphi\right)\,,\\
\dot{\lambda}\mu^{2}= & -2\mu\dot{\mu}\lambda-3H\varepsilon\,,\end{align*}
with the standard solution $\varepsilon=\varepsilon_{0}a^{-3}$. Note
that there is a degeneracy in $\varphi\left(t\right)$. In this setup
for different $\mu\left(\varphi\right)$, the same evolution of the
energy density $\varepsilon\left(t\right)$ corresponds to different
$\varphi\left(t\right)$. Again, by our discussion of cosmological
perturbations in section \ref{s:genperts}, the evolution of $\Phi$
and $\delta_{\varepsilon}$ exactly reproduces the results for the
standard dust-dominated universe.

\subsection{$\lambda\varphi$-\emph{Fluid} Possessing a Scaling Solution }

\label{s:constW}Moving beyond a constant $\mu$, let us consider
a class of models with\begin{align}
K & =\sigma X\,,\quad\quad\mbox{where} & \!\!\!\sigma=\pm1\,,\label{bla}\\
\mu & =\mu_{0}\exp\left(-\frac{\varphi}{m}\right)\,,\label{mu(phi)}\end{align}
where the mass scale for $\varphi$ is \begin{equation}
m=\sqrt{\frac{8}{3}}\,\frac{\sqrt{\sigma w_{\text{fin}}}}{1+w_{\text{fin}}}M_{\text{Pl}}\,.\label{AlphaDef}\end{equation}
In the following, we will show that, in this class of models, the
dynamics of the $\lambda\varphi$\emph{-fluid}-dominated cosmological
background has a fixed point with a constant equation of state $w_{\text{fin}}$,
where $w_{\text{fin}}$ can have either sign and can even be phantom-like.%
\footnote{Obviously we cannot realise $w_{\text{fin}}=0$ in this setup.%
}. Further, we will show that this fixed point solution is an attractor
provided $w_{\text{fin}}<1$. 

From Eqs \eqref{pressure}, \eqref{EnergyDensity} we obtain \begin{eqnarray}
p=\frac{\sigma}{2}\mu^{2} & \mbox{and} & \varepsilon=\mu^{2}\left(\frac{\sigma}{2}+\lambda\right)\,,\label{pEsimple}\end{eqnarray}
so that the \emph{instantaneous }equation of state, \begin{equation}
w_{X}=\frac{1}{1+2\sigma\lambda}\,,\label{WofLambda}\end{equation}
is determined by the value of $\lambda$. From this equation it follows
that, if $w_{X}=\mbox{const},$ then \begin{equation}
\lambda_{w_{X}}=\frac{1}{2}\sigma\left(w_{X}^{-1}-1\right)=\mbox{const\,}.\label{LambdaW}\end{equation}
In particular, for values of $w_{X}$ corresponding to an accelerating
expansion, $\lambda_{w_{X}}$ is a number of order one. Meanwhile,
the exponential form of $\mu\left(\varphi\right)$, Eq.\ (\ref{mu(phi)})
implies that the evolution of $\mu$ has a very simple form,\begin{eqnarray}
\mu=\frac{m}{t} & \mbox{and} & \mu_{\varphi}=-\frac{\mu}{m}=-\frac{1}{t}\,,\label{muEvolution}\end{eqnarray}
where we have chosen constants of integration in such a way that the
pressure $p$ of the $\lambda\varphi$-\emph{fluid} is singular exactly
at the Big Bang, $t=0$. 

We can now rewrite the equation of motion Eq.\ (\ref{CosmologyEoM})
combined with the Friedmann equation for a universe containing solely
the $\lambda\varphi$\emph{-fluid}, to obtain the equation of motion
for the equation of state\begin{equation}
w'_{X}=3w_{X}\left(1+w_{X}-\sqrt{\frac{w_{X}}{w_{\text{fin}}}}\left(1+w_{\text{fin}}\right)\right)\,.\label{LambdaEvolutionWeqConst}\end{equation}
It is easy to see that $w_{\text{fin}}$ is the fixed point of this
equation, and therefore also of the equation of motion for $\lambda$.
In the limit of $w_{\text{fin}}\rightarrow-1$, this equation reduces
to the evolution of $w$ for $\Lambda$CDM. Therefore, as $w_{\text{fin}}$
approaches that limit, the transition from matter domination to dark
energy domination becomes indistinguishable from that in $\Lambda$CDM.

Let us now consider the stability of this fixed point. Linearising
the above around $w_{X}=w_{\text{fin}}$, we obtain\[
\delta w'_{X}=\frac{3}{2}(w_{\text{fin}}-1)\delta w_{X}\,.\]
From this result, it follows that in an expanding universe, for $w_{\text{fin}}<1$,
the equation of state approaches $w_{\text{fin}}$ and $\lambda$
approaches the fixed point $\lambda_{w_{\text{fin}}}$. We call this
solution the $w$-attractor. 

Eq.\ \eqref{LambdaEvolutionWeqConst} can actually be solved, albeit
implicitly, allowing us to obtain the scale factor $a$ as a function
of the equation of state parameter $w_{X}$: \begin{equation}
\left(\frac{a}{a_{0}}\right)^{3\left(w_{\text{fin}}-1\right)}=\left[\frac{\left(\sqrt{w_{\text{fin}}/w_{X}}-w_{\text{fin}}\right)^{w_{\text{fin}}}}{\sqrt{w_{\text{fin}}/w_{X}}-1}\right]^{2}\,,\label{a(W)}\end{equation}
$ $$ $ where $a_{0}$ is a constant of integration. 

It is instructive to compare the evolution of the $\lambda\varphi$\emph{-fluid}
given by Eq.\ \eqref{a(W)} or Eq.\ \eqref{LambdaEvolutionWeqConst}
with the case of a mixture of dust and DE with a constant equation
of state $w=w_{\text{fin}}$. For $w_{\text{fin}}<0$ the evolution
of the universe has the same late-time asymptotic as in the $\lambda\varphi$\emph{-fluid}
case.\emph{ }From the\emph{ }Friedmann equations\emph{ }\eqref{Friedmann}
and \eqref{SecondFriedmann} we obtain \begin{equation}
w'_{\text{dark}}=3w_{\text{dark}}\left(w_{\text{dark}}-w_{\text{fin}}\right)\,,\label{WevolutionMixture}\end{equation}
instead of Eq.\ \eqref{LambdaEvolutionWeqConst}, where we have denoted
the total effective equation of state as $w_{\text{dark}}$. These
two equations \eqref{LambdaEvolutionWeqConst} and \eqref{WevolutionMixture}
only coincide in the limit of $w_{\text{fin}}\rightarrow-1$. For
illustrative purposes we also present the solution for \eqref{WevolutionMixture}
in the form similar to Eq.\ \eqref{a(W)}: \[
\left(\frac{a}{a_{0}}\right)^{3w_{\text{fin}}}=\frac{w_{\text{fin}}}{w_{\text{dark}}}-1\,.\]
From this analysis then, the evolution history of the $\lambda\varphi$\emph{-fluid}
cannot be reduced to the evolution of a mixture of DE with a constant
equation of state parameter $w_{\text{fin}}$ and dust, providing
a potential observational probe for such theories.

In the following, we present cosmologies in the two limits where the
$\lambda\varphi$\emph{-fluid} dominates the total energy density
of the universe and when it is subdominant to some other matter fluid.

\subsubsection{Dominant $\lambda\varphi$-Fluid in FRW Universe }

Let us again consider the model given by Eqs (\ref{pEsimple}) and
(\ref{bla}), however, this time in the case with $\lambda\gg1$,
far off the $w$-attractor. Here, the system is effectively pressureless,
by virtue of Eq.\ \eqref{WofLambda}. The energy density scales as
$a^{-3}$ since $w_{X}\approx0$. The evolution off the attractor
therefore resembles a matter-dominated epoch, eventually approaching
an era with a constant $w_{X}=w_{\text{fin}}$.

Approximating Eq.\ \eqref{LambdaEvolutionWeqConst} for small $w_{X}$
we can calculate how the equation of state evolves during this period:\[
w'_{X}\simeq3w_{X}\quad\Rightarrow\quad w_{X}\sim a^{3}\quad\Rightarrow\quad\lambda\sim a^{-3}\,.\]
Turning to perturbations during this matter-domination era, the pressure
corrections Eq.\ (\ref{GammaDEF}) can be written as \begin{equation}
\Gamma\left(w_{X}\right)=3w_{X}\sqrt{\frac{w_{X}}{w_{\text{fin}}}}\left(\frac{1+w_{\text{fin}}}{1+w_{X}}\right)\sim a^{-9/2}\,.\label{GammaOurModel}\end{equation}
where we can explicitly see that $\Gamma\rightarrow0$ as $w_{\text{fin}}\rightarrow-1$.
For parameter values motivated by dark energy, $\left|1+w_{\text{fin}}\right|\ll1$,
these pressure corrections to the evolution of the potential $\Phi$
are subleading until $w_{X}$ approaches its attractor value, since
in Eq.\ (\ref{PhiN}) \begin{equation}
\Phi''+\Phi'\left(\frac{5}{2}-\frac{3}{2}w_{X}+\Gamma\right)+\left(-3w_{X}+\Gamma\right)\Phi=0\,.\label{PhiN_wX}\end{equation}
So, while off the attractor, there is no significant deviation in
the growth of structure from standard considerations. 

On the attractor, $w_{X}=w_{\text{fin}}$, we obtain \[
\Gamma=3w_{\text{fin}}\,,\]
and Eq.\ \eqref{PhiN_wX} confirms what we found in section \ref{s:genperts}
for the general model: for $w_{\text{fin}}>-5/3$, $\Phi=\text{const}$,
while $\delta_{\varepsilon}\sim a^{1+3w_{\text{fin}}}$ is the growing
mode inside the horizon.

\subsubsection{Subdominant $\lambda\varphi$-Fluid in FRW Universe}

Motivated by the discussion above let us still consider theories with
\[
\mu=\mu_{0}\exp\left(-\frac{\varphi}{m}\right)\,,\]
where $m=\mbox{const}$ and may, for example, be given by Eq.\ (\ref{AlphaDef}).
Using the result of Eq.\ (\ref{muEvolution}) we obtain the equation
of motion for $\lambda$, \begin{equation}
\dot{\lambda}=\frac{1}{t}\left(\sigma+2\lambda\right)-3H\left(\sigma+\lambda\right)\,.\label{LambdaSubdominant}\end{equation}
In the history of the universe there were at least two stages with
the background equation of state, $w_{\text{b}}$, approximately constant,
namely the radiation-dominated epoch---with $w_{\text{b}}\simeq\frac{1}{3}$---and
the matter-dominated era---with $w_{\text{b}}\simeq0$. Let us then
consider the dynamics of Eq.\ (\ref{LambdaSubdominant}) when the matter
content of universe consists mostly of some fluid with a constant
equation of state parameter, $w_{\text{b}}=\mbox{const}$. In that
case, we have\begin{eqnarray*}
a=\left(\frac{t}{t_{0}}\right)^{2/3\left(w_{\text{b}}+1\right)} & \mbox{and consequently} & H=\frac{2}{3\left(w_{\text{b}}+1\right)}\frac{1}{t}\,,\end{eqnarray*}
here $t_{0}$ would correspond to the age of the universe today, when
$a=1$, if $w_{\text{b}}=\mbox{const}$. On this background, Eq.\ \eqref{LambdaSubdominant}
for the subdominant $\lambda\varphi$-\emph{fluid }can be expressed
as\begin{equation}
\frac{\mathrm{d}\lambda}{\mathrm{d}\ln t}=\sigma\frac{w_{\text{b}}-1}{w_{\text{b}}+1}+\frac{2w_{\text{b}}}{w_{\text{b}}+1}\lambda\,.\label{LambdaSubdominantWconst}\end{equation}
This equation has the following solution\begin{align*}
\lambda\left(t\right)= & \lambda_{w_{\text{b}}}+Ct^{2w_{\text{b}}/\left(w_{\text{b}}+1\right)}=\lambda_{w_{\text{b}}}+Ca^{3w_{\text{b}}}, & \mbox{for}\: & w_{\text{b}}\neq0\,,\\
\lambda\left(t\right)= & -\ln\left(\frac{t}{t_{1}}\right), & \mbox{for}\: & w_{\text{b}}=0\,,\end{align*}
where $\lambda_{w}$ is the fixed point solution given by the formula
(\ref{LambdaW}) and $C$ and $t_{1}$ are constants of integration.
We can therefore see that, for $w_{\text{b}}\neq0$, the $\lambda\varphi$-\emph{fluid
}corresponds to a \emph{mixture }of a fluid with the same equation
of state $w_{\text{b}}$ as the background and, in addition, dust
\begin{align*}
\mbox{\ensuremath{p=}} & \frac{\sigma}{2}m^{2}t^{-2}=\frac{\sigma}{2}\left(\frac{m}{t_{0}}\right)^{2}a^{-3\left(1+w_{\text{b}}\right)}\,,\\
\varepsilon= & w_{\text{b}}^{-1}p+\left(\frac{m}{t_{0}}\right)^{2}Ca^{-3}\,,\end{align*}
while for a dust-like background, $w_{\text{b}}=0$, the $\lambda\varphi$-\emph{fluid}'s\emph{
}hydrodynamics obeys\begin{align*}
p= & \frac{\sigma}{2}m^{2}t^{-2}=\frac{\sigma}{2}\left(\frac{m}{t_{0}}\right)^{2}a^{-3}\,,\\
\varepsilon= & p-\frac{m^{2}}{t^{2}}\ln\left(\frac{t}{t_{1}}\right)=p\left(1-2\sigma\ln\left(\frac{t}{t_{1}}\right)\right)\,.\end{align*}
In the late-time asymptotic, the equation of state of the $\lambda\varphi$\emph{-fluid
}approaches that of the background provided that $w_{\text{b}}<0.$
In particular, this tracking behaviour means that in an inflationary
background when $w_{\text{b}}\simeq-1$, the $\lambda\varphi$\emph{-fluid}
does not redshift away but instead survives in form of an effective
cosmological constant.

\section{Dusty Dark Energy}

\label{s:DDE}In this section, we will present a unified dark matter
and dark energy model using a single dynamical degree of freedom,
which we call \emph{Dusty Dark Energy}.

The discussion of the constant-$w$ model presented in section \ref{s:constW}
is suggestive: a model with $\mu(\varphi)$ of the form given in Eq.
\eqref{mu(phi)} has a background evolution that evolves as dust when
it is far off its attractor and eventually settles on a constant equation
of state determined by the value of the mass-scale parameter $m$.
This picture is not altered when the $\lambda\varphi$\emph{-fluid}
is subdominant during a radiation-domination epoch: the dust-like
component will eventually overwhelm the redshifting radiation resulting
in a period of expansion similar to matter domination with its duration
determined by the value of $\lambda$ when radiation becomes subdominant.
Once $\lambda$ approaches a number of order unity, the transition
to dark-energy domination will occur at a rate similar to the usual
transition in $\Lambda$CDM and eventually the expansion will settle
on a constant (and arbitrary) equation of state. This background evolution
provides an expansion history that is close to $\Lambda$CDM and yet
does not explicitly contain separate dark-matter and dark-energy components:
only radiation and our $\lambda\varphi$\emph{-fluid }are necessary.

Beyond describing the background dynamics, we have to show that the
perturbations in the fluid evolve similarly to those in $\Lambda$CDM.
This is necessary not only for the growth of large-scale structure
during matter domination, but also during the radiation-domination
epoch: the anisotropies in the cosmic microwave background depend
on the existence of appropriate gravitational potentials driven by
dark matter.

In section \ref{s:constW}, we have shown that during matter domination
and at late times perturbations evolve similarly to the CDM perturbations
in $\Lambda$CDM. The remaining piece is to prove that the perturbations
evolve in a dust-like manner even when the energy density is dominated
by radiation, which we will show as follows.

During domination by radiation---or any external fluid in general---the
Newtonian potential, $\Phi$, is driven by perturbations in that fluid.
The subdominant components then respond to this potential and evolve
according to equations of motion arising from the conservation of
their EMT. For dust, the standard result, written in a suggestive
manner, is\begin{align}
 & \left(\frac{\delta\varphi_{\text{dust}}}{\mu}\right)^{\cdot}=\Phi\,,\\
 & \dot{\delta}_{\text{dust}}=3\dot{\Phi}+\frac{\Delta}{a^{2}}\left(\frac{\delta\varphi_{\text{dust}}}{\mu}\right)\,.\nonumber \end{align}
In the case of a subdominant Dusty Dark Energy, the equivalent equations,
Eq.\ \eqref{PerturbtLambdaEOM} and \eqref{dotRelativeDeltaE} tell
us that the eventual evolution to a matter-dominated era requires
that $\lambda\gg1$, implying $w_{X}\sim1/2\lambda$. Therefore we
can approximate\begin{eqnarray}
 &  & \left(\frac{\delta\varphi}{\mu}\right)^{.}=\Phi\,,\label{PertCons-1}\\
 &  & \dot{\delta}_{\varepsilon}\simeq3\dot{\Phi}+\frac{\Delta}{a^{2}}\left(\frac{\delta\varphi}{\mu}\right)-\nonumber \\
 &  & \quad\quad\quad-3Hm^{-1}\lambda^{-1}\delta\varphi-\frac{3}{2}H\lambda^{-1}\delta_{\varepsilon}\,.\label{RelativedeltaEinRadiat-1}\end{eqnarray}
All we now need to do is to show that the terms additional to the
equations for CDM are negligible. For $\lambda\gg1$, \[
\dot{\delta}_{\varepsilon}\sim H\delta_{\varepsilon}\gg H\lambda^{-1}\delta_{\varepsilon}\,.\]
From Eq.\ \eqref{PertCons-1} we can obtain the estimate \[
\delta\varphi\sim\frac{\mu\Phi}{H}\,.\]
Then, the pressure correction in Eq.\ (\ref{RelativedeltaEinRadiat-1})
is small compared to the Newtonian potential providing \[
\dot{\Phi}\gtrsim\Phi H\gg\mu\Phi m^{-1}\lambda^{-1}\,,\]
or, in other words, if \begin{equation}
1\gg\frac{\mu}{mH\lambda}\,.\label{condition}\end{equation}
On the other hand \[
\frac{\mu}{mH}\sim\frac{\left(1+w_{\text{fin}}\right)\sqrt{-p_{X}}}{HM_{\text{Pl}}}\sim\left(1+w_{\text{fin}}\right)\sqrt{\frac{-p_{X}}{p_{\text{rad}}}}\,.\]
Since we assume that the DDE is subdominant, $\varepsilon_{\text{rad}}=3p_{\text{rad}}\gg\left|p_{X}\right|$
and, by construction, $\left|1+w_{\text{fin}}\right|\ll1$. Therefore
all the additional terms in the evolution equations are negligible,
and the perturbations in the DDE obey the same equations for evolution
as standard dust during radiation domination, demonstrating that the
DDE model provides for a viable cosmology with our fluid playing the
role of \emph{both }cold dark matter and dark energy.

The duration of the matter-domination era is determined by the initial
values of $\lambda(t_{0})\gg1$ for any given initial $\varphi(t_{0})$.
The former determines the time when the DDE turns over from dark-matter-like
to dark-energy-like, while the latter sets the time where it begins
to dominate over radiation. Since we have to tune both values, we
have not provided a solution to the coincidence problem. On the other
hand, this model is a ``minimalist'' description---as we require only
two initial conditions for set the transitions of the two epochs.

\section{Summary}

In this paper, we introduced a novel class of field theories with
a single dynamical degree of freedom which have a perfect-fluid interpretation.
The key feature of this theory is that its fluid velocity flows along
geodesics---hence mimicking ``dust'' in this respect. On the other
hand, unlike a standard cold-dark-matter fluid, it carries pressure
parallel to its fluid velocity. In cosmology, this pressure affects
the expansion history.

This sleight-of-hand is achieved by means of a ``Lagrange multiplier''
field, employed to constrain by equation of motion the above-mentioned
behaviour of the fluid velocity vector. Our system then consists of
\emph{two first order equations of motion}, and hence effectively
a single degree of freedom. This dynamic cannot be reproduced by usual
scalar field theories such as k-\emph{essence} or higher derivative
theories.

As an application, we consider the evolution and effects of this fluid
in cosmology. We show that there exists a class of scaling solutions
which have an attractor solution with fixed equation of state $w_{\mathrm{fin}}$.
Off the attractor, this model possess an interesting dynamic where
part of the energy density redshifts as dust while part of the energy
density tracks any dominant background energy density. We use this
curious property to construct a unified dark energy/dark matter model
where the limit $w_{\mathrm{fin}}=-1$ corresponds to standard $\Lambda$CDM.

We also show that in this class of models, we can construct phantom
models with $w_{\mathrm{fin}}<-1$ where there is no pathology when
crossing the ``phantom divide'' at $w=-1$, at least classically.
If we insist that the system satisfy the Null Energy Condition (i.e.
$w\geq-1$), then we show that the kinetic term for perturbations
is positive definite. 

Finally, we would like to conclude by emphasising that this class
of theories provides a novel framework for cosmological model building
and exploring exotic states of matter.

\medskip{}

\begin{acknowledgments}
It is a pleasure to thank Gregory Gabadadze, Andrei Gruzinov, Lam
Hui, Dan Kabat, Justin Khoury, Dmitry Malyshev, Marc Manera, Slava
Mukhanov, Alberto Nicolis, Oriol Pujol\`as, Sergei Sibiryakov and
Arkady Vainshtein for very useful discussions and criticism. We thank Paolo Creminelli for fruitful correspondence regarding the possible equivalence of our model and quintessence with zero sound speed. I.$\,$S.
and A.$\,$V. are grateful to the organisers and staff of the 45$^{\text{th}}$
Rencontres de Moriond, Cosmology Session and the Metropolitan Transportation
Authority of New York for their hospitality during the final stages
of preparation of this paper. The work of I.$\,$S. and A.$\,$V.
was supported by the James Arthur Fellowship.
\end{acknowledgments}

\appendix
\section{\label{Appendix}Derivation of perturbed energy conservation}

The differential form of the perturbed constraint equation (\ref{PerturbtLambdaEOM})
is \begin{equation}
\delta\ddot{\varphi}=\left(\mu_{\varphi}^{2}+\mu\mu_{\varphi\varphi}\right)\delta\varphi+\dot{\Phi}\mu+2\Phi\mu\mu_{\varphi}\,.\label{DifferentialPerturbedConstraint}\end{equation}
Let us perturb the energy-conservation equation (\ref{EnergyConservation}):
\begin{equation}
\delta u^{\mu}\nabla_{\mu}\varepsilon+\delta\dot{\varepsilon}+\delta\theta\left(\varepsilon+p\right)+3H\left(\delta\varepsilon+\delta p\right)=0\,.\label{perturEnergyCons}\end{equation}
First of all, as usual we have $\delta u^{t}=-\Phi$ so that \begin{equation}
\delta u^{\mu}\nabla_{\mu}\varepsilon=\dot{\varepsilon}\delta u^{t}=-\Phi\dot{\varepsilon}\,.\label{deltaUgradE}\end{equation}
Further, we perturb the formula for the expansion Eq.\ (\ref{expansion})
\[
\delta\theta=\mu^{-1}\delta\left(\Box\varphi\right)-\left(\mu^{-2}\Box\varphi\mu_{\varphi}+\mu_{\varphi\varphi}\right)\delta\varphi\,.\]
For the perturbations of the d'Alambertian we have the standard result
\[
\delta\left(\Box\varphi\right)=-4\dot{\Phi}\dot{\varphi}-2\Phi\ddot{\varphi}-6H\Phi\dot{\varphi}+\delta\ddot{\varphi}+3H\delta\dot{\varphi}-\frac{\Delta}{a^{2}}\delta\varphi\,,\]
which after using Eqs (\ref{dotPhi}) and (\ref{PerturbtLambdaEOM})
can be written in terms of $\Phi$ and $\delta\varphi$ as \[
\delta\left(\Box\varphi\right)=\left(\mu_{\varphi}^{2}+\mu\mu_{\varphi\varphi}+3H\mu_{\varphi}-\frac{\Delta}{a^{2}}\right)\delta\varphi-3\mu\left(\dot{\Phi}+H\Phi\right)\,.\]
Therefore for the perturbation of the expansion we obtain \[
\delta\theta=-3\left(\dot{\Phi}+H\Phi\right)-\mu^{-1}\frac{\Delta}{a^{2}}\delta\varphi\,.\]
Substituting this expression along with along the formula (\ref{deltaUgradE})
into Eq.\ (\ref{perturEnergyCons}) results in \[
\delta\dot{\varepsilon}-\left(3\dot{\Phi}+\frac{\Delta}{a^{2}}\left(\frac{\delta\varphi}{\mu}\right)\right)\left(\varepsilon+p\right)+3H\left(\delta\varepsilon+\delta p\right)=0\,,\]
which corresponds to the standard case (c.f.\ \cite[p. 312, Eq.\ (7.105)]{Mukhanov:2005sc}). 

\bibliographystyle{utphys}
\bibliography{ListDustA}

\end{document}